\documentstyle[11pt,epsfig,twoside,cite,amssymb]{article}
\newcommand{\bc}{\begin{center}}
\newcommand{\ec}{\end{center}}
\begin{document}
\begin{flushright}
PRA-HEP 00-03
\end{flushright}
\begin{center}
{\Large \bf QCD improved parton distribution functions of
${\mathbf {\gamma^*_L}}$}

\vspace*{0.3cm}
Ji\v{r}\'{\i} Ch\'{y}la

\vspace*{0.2cm}
{\em Institute of Physics, Na Slovance 2, Prague 8, Czech Republic}

\vspace*{0.5cm}
{\large \bf Abstract}
\begin{quotation}
\noindent
QCD corrections to the QED formula for parton distribution functions
of the longitudinal virtual photon are derived in the
leading--logarithmic approximation. It is shown that the resulting
PDF satisfy the same homogeneous evolution equations as
those of hadrons, but contrary to the latter are perturbatively
calculable. Properties of these distribution functions are discussed
and their phenomenological consequences outlined.
\end{quotation}
\end{center}

\section{Introduction}
Parton distribution functions (PDF) of the photon
can be written as sums of the
{\em hadronic}
(HAD) and {\em pointlike}
\footnote{Frequently also called {\em anomalous}, because they
contain the term that is absent in the case of hadrons. I prefer
the denomination {\em pointlike}, which reflects the fact that
they arise from pointlike $\gamma\rightarrow q\overline{q}$
coupling.}
(PL) parts. The former satisfy the same homogeneous evolution
equations as those of hadrons, whereas the latter satisfy the
inhomogeneous ones. Within a subset of pointlike solutions specified
by the value of the scale $M_0$ at which they vanish we can thus
write
\begin{equation}
D(x,M_0,M)= D^{\mathrm {HAD}}(x,M_0,M)+D^{\mathrm{PL}}(x,M_0,M),
~~~~~D=q,\overline{q},G.
\label{separation}
\end{equation}
Due to the freedom in the choice of $M_0$ the separation of PDF
into their pointlike and hadronic parts is, however, ambiguous. In
\cite{smarkem1} we discussed practical aspects of this ambiguity for
the Schuler--Sj\"{o}strand sets of parameterizations \cite{sas1}.
The quark distribution functions of the photon can be
written as combinations of singlet and nonsinglet parts
\begin{equation}
\Sigma \equiv \sum_{i=1}^{n_f}\left(q_i+\overline{q}_i\right),~~~
q_i^{\mathrm{NS}} \equiv  \left(q_i+\overline{q}_i\right)-
\frac{1}{n_f}\Sigma.
\label{sinsi}
\end{equation}
Due to the presence of hadronic contributions, individual
$q^{\mathrm{NS}}_i$ are in general independent nonsinglet
distributions. Their pointlike parts are, however,
proportional to the corresponding charges and (neglecting quark mass
effects) pointlike parts of $q_i$ can therefore be expressed in
terms of one singlet and one nonsinglet quantity
\begin{equation}
q_{i}^{\mathrm{PL}}=
\left(e_i^2-\langle e^2\rangle\right)q^{\mathrm{NS,PL}}+
\langle e^2\rangle q^{\mathrm{{S,PL}}},~~
\Sigma =2n_f\langle e^2\rangle q^{\mathrm{S,PL}}.
\label{decomposition}
\end{equation}
The pointlike part of $q^{\mathrm{NS}}$ is defined via the
resummation of multiple gluon emissions off the quark from the
primary $\gamma\rightarrow q\overline{q}$ splitting. In units of
$\alpha/2\pi$
\footnote{Throughout this paper the expressions for PDF of the
photon will be given in these units.}
\begin{displaymath}
q^{\mathrm{NS,PL}}(x,M_0,M) \equiv
k_{\mathrm{NS}}(x)
\int^{M^2}_{M_0^2}\frac{{\mathrm d}\tau_1}{\tau_1}+
\int^{1}_{x}\frac{{\mathrm d}y}{y}P^{(0)}_{qq}
\left(\frac{x}{y}\right) \int^{M^2}_{M_0^2} \frac{{\mathrm
d}\tau_2}{\tau_2}\frac{\alpha_s(\tau_2)}{2\pi}
k_{\mathrm{NS}}(y)\int^{\tau_2}_{M_0^2} \frac{{\mathrm
d}\tau_1}{\tau_1}+
\end{displaymath}
\begin{equation}
\int^{1}_{x}\frac{{\mathrm d}y}{y}P^{(0)}_{qq}
\left(\frac{x}{y}\right) \int^{1}_{y}\frac{{\mathrm
d}w}{w}P^{(0)}_{qq} \left(\frac{y}{w}\right) \int^{M^2}_{M_0^2}
\frac{{\mathrm d}\tau_3}{\tau_3}\frac{\alpha_s(\tau_3)}{2\pi}
\int^{\tau_3}_{M_0^2} \frac{{\mathrm
d}\tau_2}{\tau_2}\frac{\alpha_s(\tau_2)}{2\pi}
k_{\mathrm{NS}}(w)\int^{\tau_2}_{M_0^2}\frac{{\mathrm
d}\tau_1}{\tau_1} +\cdots,
\label{Tresummation}
\end{equation}
where $k_{\mathrm{NS}}=3(x^2+(1-x)^2) $ and $\alpha_s(M)$
satisfies the standard renormalization group equation
\begin{equation}
\frac{{\mathrm d}\alpha_s(\mu)}{{\mathrm d}\ln \mu^2}\equiv
\beta(\alpha_s(\mu))=
-\frac{\beta_0}{4\pi}\alpha_s^2(\mu)-
\frac{\beta_1}{16\pi^2}
\alpha_s^3(\mu)+\cdots,
\label{RG}
\end{equation}
where, in QCD with $n_f$ massless quark flavours, the first two
coefficients, $\beta_0=11-2n_f/3$ and $\beta_1=102-38n_f/3$, are unique,
while all the higher order ones are ambiguous. In this paper only the
LO QCD corrections will be considered and therefore only the first term
in (\ref{RG}) will be taken into account. In terms of standard moments
the series in (\ref{Tresummation}) resums to
\begin{equation}
q^{\mathrm {NS,PL}}(n,M_0,M)=\frac{4\pi}{\alpha_s(M)}
\left[1-\left(\frac{\alpha_s(M)}{\alpha_s(M_0)}\right)^
{1-2P^{(0)}_{qq}(n)/\beta_0}\right]a_{\mathrm {NS}}(n),~~~
a_{\mathrm {NS}}(n)\equiv
\frac{k^{(0)}(n)}{\beta_0-2P^{(0)}_{qq}(n)}.
\label{generalpointlike}
\end{equation}
For transverse virtual photon $\gamma_T^*$ the above considerations
can be repeated with only a minor modification reflecting the fact that
its momentum $P_{\gamma}$ is off-shell, $P^2\equiv -P_{\gamma}^2>0$.
The exact meaning of the quantity $\tau$ denoting in
(\ref{Tresummation}) the outgoing parton off-shellness is
somewhat ambiguous. I have taken
$\tau$ as $\tau\equiv -(p^2-m^2)/x$, where $p$ and $m_q$ stand for
parton momentum and mass respectively. Working out the kinematics of
the $\gamma^*\rightarrow q\overline{q}$ vertex in the
collinear limit yields $\tau^{\mathrm{min}}=P^2+m_q^2/(x(1-x))$.
For massless quarks this implies replacing in
(\ref{Tresummation}) $M_0^2$ with $P^2$.

\section{The concept of PDF of $\gamma_L^*$ in QED}
Before discussing PDF of $\gamma_L^*$ in the framework of QCD,
let me illustrate the usefulness of this concept on the QED analysis
of DIS of electrons on virtual photons. In the region
$m_q^2\ll P^2\ll Q^2$ experiments measure the following quantity
\begin{equation}
F_{\mathrm{eff}}^{\gamma}(x,P^2,Q^2)\equiv \frac{Q^2}{4\pi^2\alpha}
\left(\sigma_{TT}+\sigma_{LT}+
\sigma_{TL}+\sigma_{LL}\right)
=\frac{Q^2}{4\pi^2\alpha}\sigma(P^2,Q^2,W^2),
\label{Feff}
\end{equation}
where the cross--sections
\footnote{The first and second indices corresponding to
probe and target photon respectively.}
$\sigma_{jk}$ are in general functions of $W^2,P^2$
and $Q^2$ and $x=Q^2/(W^2+Q^2+P^2)$. The ${\cal O}(\alpha)$ QED
expressions for individual contributions to $F_{\mathrm{eff}}^{\gamma}$
read \cite{russians}
\begin{eqnarray}
F_{TT}^{\mathrm{QED}}(x,P^2,Q^2)& = & 6\,\biggl[\,
\underbrace{\left(x^2+(1-x)^2\right)\ln\frac{Q^2}{x^2P^2}+
(2x(1-x)-1)}
_{\mathrm{partonic}}+\underbrace{(2x(1-x)-1)}_{\mathrm{nonpartonic}}\,
\biggr],
\label{TT}\\
F_{LT}^{\mathrm{QED}}(x,P^2,Q^2)& = &6\,[\,\underbrace{4x(1-x)}_
{\mathrm{nonpartonic}}\,],\label{LT}\\
F_{TL}^{\mathrm{QED}}(x,P^2,Q^2) & = & 6\,[\,\underbrace{4x(1-x)}_
{\mathrm{partonic}}\,],~~~~~~~~F_{LL}^{\mathrm{QED}}(x,P^2,Q^2)=0.
\label{TL}
\end{eqnarray}
The denomination ``partonic''
refers to the fact that the corresponding term
arises from integration over $\tau$ of the quark
from the primary splitting $\gamma\rightarrow q\overline{q}$ close to
its lower limit $\tau^{\mathrm{min}}=P^2$, i.e. from
the region of phase space where the emitted quarks (antiquarks) are
almost collinear with the incoming photon. In the case of
$F_{TT}^{\mathrm{QED}}$ part of the integrand is proportional to $1/\tau$,
which yields the logarithmic term, whereas the ``partonic''
$2x(1-x)-1=-(x^2+(1-x)^2)$ in $F_{TT}^{\mathrm{QED}}$ and $4x(1-x)$
in $F_{TL}^{\mathrm{QED}}$ arise from integration over $P^2/\tau^2$. In
both cases the resulting expressions have a clear parton model
interpretation. On the other hand, the ``nonpartonic'' contributions
result from integration over the whole phase space and cannot be
interpreted in terms of PDF. Adding all terms in (\ref{Feff}) we get
\begin{equation}
F_{\mathrm{eff}}^{\mathrm{QED}}(x,P^2,Q^2)=6\left[
\left(x^2+(1-x)^2\right)\ln\frac{Q^2}{x^2P^2}+12x(1-x)-2\right].
\label{all}
\end{equation}
One might argue that there is no reason to distinguish the origins and
interpretation of the individual terms that add up to the
nonlogarithmic part of (\ref{all}), and consequently, the effects of
target $\gamma_L^*$ can be included as part of the ``constant'' term in
(\ref{all}). Although legitimate, I prefer not to adopt this procedure
and keep track of the origins and interpretation of individual
contributions to the constant part of $F_{\mathrm{eff}}^{\gamma}$
because:
\begin{itemize}
\item The ``partonic'' terms, coming from the vicinity of
mass singularities are {\em universal}
quantities, characterizing the incoming $\gamma_L^*$, whereas the
``nonpartonic'' ones are process dependent.
\item Virtuality dependence of different terms is different.
The above expressions hold for $m_q^2\ll P^2\ll Q^2$. As
$P^2/m_q^2\rightarrow 0$ the nonpartonic terms, coming from
integration over the whole phase space, remain unchanged, whereas
$F_{TL}^{\gamma}$ vanishes by gauge invariance as
$F_{TL}^{\gamma}\propto P^2/m_q^2$. In the case of
$F_{TT}^{\mathrm{QED}}$ the partonic part of the constant
term gets additional contribution from the term
$m_q^2/(m_q^2+x(1-x)P^2)\rightarrow 1$, absent for
$P^2\gg m_q^2$. Taking these facts into account allows us to
derive in a physically transparent way the constant term of
$F^{\mathrm{QED}}_{\mathrm{eff}}(x,Q^2)$ for the real photon:
$[2x(1-x)-1]+[2x(1-x)-1]+4x(1-x)+1=8x(1-x)-1$.
\end{itemize}
The fact that $F_{TL}^{\mathrm{QED}}=24x(1-x)$, which does have a
parton model interpretation coincides with $F_{LT}^{\mathrm{QED}}$,
which does not, reflects the symmetry of the total cross
section $\sigma_{TL}(\gamma^*(Q^2)\gamma^*(P^2))$ with respect to
$P^2$ and $Q^2$.

\section{Pointlike part of $\gamma_L^*$ in QCD -- nonsinglet
distribution}
\begin{figure}[t]\centering
\epsfig{file=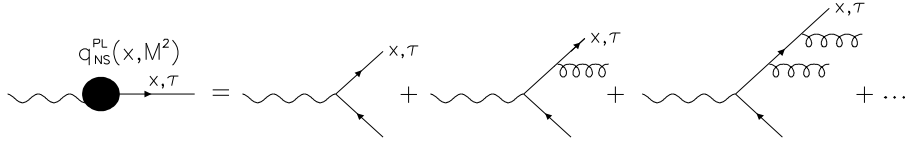,width=12cm}
\caption{Diagrams defining the pointlike part of nonsinglet
quark distribution function of the photon. The resummation involves
integration over quark virtualities $\tau\le M^2$.}
\label{figpl}
\end{figure}

\begin{figure}[t]\centering
\epsfig{file=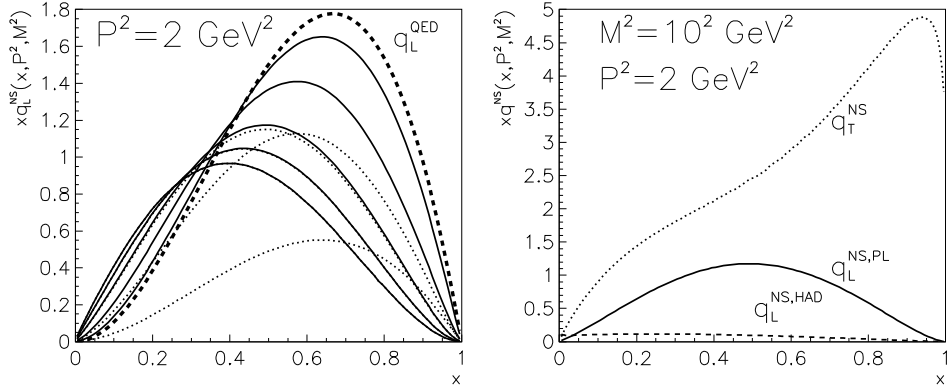,width=13cm}
\caption{Left: Scaling violations for $q_L^{\mathrm{NS}}$. Solid
curves correspond to (\ref{result}) for $P^2=2$ GeV$^2$ and
(from above at large $x$) $M^2=3,10,10^2,10^3,10^4$ GeV$^2$, the dotted
ones to the same values of $M^2$ but the formula (\ref{result2}).
For the highest two values of $M^2$ the dotted curves are
indistinguishable from the solid ones.
Right: Comparison of the pointlike part of $q_L^{\mathrm{NS}}$ as given
by (\ref{result}) with the hadronic part of $q_T^{\mathrm{NS}}$
evaluated from SAS2D parameterization of $u_T$ according to the
formula (\ref{rescaled}).}
\label{scaling}
\end{figure}

The purely QED prediction for the pointlike part
\footnote{In not stated otherwise all expressions for quark and gluon
distribution functions mentioned in the following concern their
pointlike part and the superscript ``PL'' will therefore be dropped.}
of $q_L^{\mathrm{NS}}$ depends on the quark mass which provides the
scale governing its threshold behavior. In this paper I consider QCD
corrections to QED formula for massless quarks
(or for $x(1-x)P^2\gg m_q^2$) and in the region $P^2\ll M^2$.
For $\gamma_L^*$ the resummation of diagrams in Fig. \ref{figpl}
in the collinear region takes the form
\begin{displaymath}
q^{\mathrm{NS}}_L(x,P^2,M^2) \equiv k_L(x)
\int^{M^2}_{P^2}{\mathrm d}\tau_1\frac{P^2}{\tau_1^2}+
\int^{1}_{x}\frac{{\mathrm d}y}{y}P^{(0)}_{qq}
\left(\frac{x}{y}\right) \int^{M^2}_{P^2} \frac{{\mathrm
d}\tau_2}{\tau_2}\frac{\alpha_s(\tau_2)}{2\pi}
k_L(y)\int^{\tau_2}_{P^2} {\mathrm d}\tau_1\frac{P^2}{\tau_1^2}+
\end{displaymath}
\begin{equation}
\int^{1}_{x}\frac{{\mathrm d}y}{y}P^{(0)}_{qq}
\left(\frac{x}{y}\right) \int^{1}_{y}\frac{{\mathrm
d}w}{w}P^{(0)}_{qq} \left(\frac{y}{w}\right) \int^{M^2}_{P^2}
\frac{{\mathrm d}\tau_3}{\tau_3}\frac{\alpha_s(\tau_3)}{2\pi}
\int^{\tau_3}_{P^2} \frac{{\mathrm
d}\tau_2}{\tau_2}\frac{\alpha_s(\tau_2)}{2\pi}
k_L(w)\int^{\tau_2}_{P^2}{\mathrm d}\tau_1\frac{P^2}{\tau_1^2} +\cdots,
\label{Lresummation}
\end{equation}
where $k_L(x)\equiv 12x(1-x)$ and the first term in
(\ref{Lresummation}) defines the QED contribution
\begin{equation}
q_L^{\mathrm {QED}}(x,P^2,M^2)=k_L(x)P^2\left(\frac{1}
{\tau_{\mathrm{min}}}-\frac{1}{M^2}\right)=k_L(x)
\left(1-\frac{P^2}{M^2}\right)
\begin{array}[t]{c}
\longrightarrow \\[-0.25cm]
{\scriptstyle P^2\rightarrow 0}
\end{array} k_L(x).
\label{QEDL}
\end{equation}
Performing in (\ref{Lresummation}) the inner integrals over $\tau_1$ we
get
\begin{eqnarray}
\lefteqn{q^{\mathrm{NS}}_L(x,P^2,M) =
k_L(x)\left(1-\frac{P^2}{M^2}\right)
+\int^1_x\frac{{\mathrm d}y}{y}P_{qq}^{(0)}\left(\frac{x}{y}\right)
k_L(y)\int^{M^2}_{P^2}\frac{{\mathrm d}\tau_2}{\tau_2}
\frac{\alpha_s(\tau_2)}{2\pi}\left(1-\frac{P^2}{\tau_2}\right)
+} & & \label{integrals}\\
& &
\int^1_w\frac{{\mathrm d}w}{w}P^{(0)}_{qq}\left(\frac{x}{w}\right)
\int_w^1 \frac{{\mathrm d}y}{y}P^{(0)}_{qq}\left(\frac{w}{y}\right)
k_L(y)\int^{M^2}_{P^2}\frac{\mathrm{d}\tau_3}{\tau_3}
\frac{\alpha_s(\tau_3)}{2\pi}
\int^{\tau_3}_{P^2}\frac{\mathrm{d}\tau_2}{\tau_2}
\frac{\alpha_s(\tau_2)}{2\pi}
\left(1-\frac{P^2}{\tau_2}\right)+\cdots
\label{integrals2}\nonumber
\end{eqnarray}
Note that $(1-P^2/M^2)$ and $(1-P^2/\tau_2)$ are defined also outside
the region $P^2\le M^2$ (or $P^2\le \tau_2$), but are negative there.
In order to avoid this unphysical region, all integrals
in (\ref{integrals}) will be understood to be multiplied by
$\theta(1-P^2/M^2)$ (or $\theta(1-P^2/\tau_2)$), guaranteeing thus
the vanishing of (\ref{integrals}) as $P^2\rightarrow M^2$.
Taking into account in each brackets of (\ref{integrals}) only the
unity, we get
\begin{eqnarray}
q_L^{\mathrm{NS}}(x,P^2,M^2)=
k_L(x)\!\!\!&+&\!\!\!\frac{2}{\beta_0}\int^1_x\frac{{\mathrm d}y}{y}
P_{qq}^{(0)}\left(\frac{x}{y}\right)k_L(y)\ln s  \label{l2}\nonumber \\
&+&\!\!\!\frac{1}{2}\left(\frac{2}{\beta_0}\right)^2
\int^1_x\frac{{\mathrm d}w}{w}
P_{qq}^{(0)}\left(\frac{x}{w}\right)\int^1_w\frac{{\mathrm d}y}{y}
P^{(0)}_{qq}\left(\frac{w}{y}\right)k_L(y)\ln^2 s+\cdots,
\label{inte}
\end{eqnarray}
where $s\equiv \ln(M^2/\Lambda^2)/\ln(P^2/\Lambda^2)$. In momentum
space (\ref{inte}) resums to
\begin{equation}
q_L^{\mathrm{NS}}(n,P^2,M^2)=
k_L(n)\left[\frac{\alpha_s(M^2)}{\alpha_s(P^2)}\right]
^{-2P^{(0)}_{qq}(n)/\beta_0},
\label{result}
\end{equation}
which satisfies standard hadronic evolution equation in the
nonsinglet channel
\begin{equation}
\frac{{\mathrm{d}}q_L^{\mathrm{NS}}(n,P^2,M^2)}{{\mathrm{d}}\ln M^2}=
\frac{\alpha_s(M^2)}{2\pi}P^{(0)}_{qq}(n)q_L^{\mathrm{NS}}(n,P^2,M^2).
\label{NSevolution}
\end{equation}
The resummation in (\ref{integrals}) resembles the definition
of nonsinglet quark distribution function of hadrons, with
$k_L(x)$ playing the role of ``bare'' quark distribution
function and $P^2$ the initial scale. Note, however, that
(\ref{result}) has been derived under the assumption $P^2\ll M^2$.
Translating (\ref{result}) into $x$-space by means of inverse Mellin
transformation leads to scaling violations illustrated by solid curves
in Fig. \ref{scaling}. Multiple gluon emission leads to softening of the
distribution with increasing $M^2$, a feature typical for hadrons.

As emphasized above, the result (\ref{result}) holds for $P^2\ll M^2$
only. For $P^2\rightarrow M^2$ the expression (\ref{integrals})
vanishes by definition, whereas (\ref{result}) goes to $k_L$.
Anticipating the results of the next
paragraph let me approximate the inclusions of the terms neglected in
(\ref{result}) by replacing the first term on the r.h.s. of (\ref{inte})
by $k_L(x)(1-P^2/M^2)$. This amounts to replacing (\ref{result}) by
\begin{equation}
q_L^{\mathrm{NS}}(n,P^2,M^2)=k_L(n)
\left(\left[\frac{\alpha_s(M^2)}{\alpha_s(P^2)}\right]
^{-2P^{(0)}_{qq}(n)/\beta_0}-\frac{P^2}{M^2}\right)
\label{result2}
\end{equation}
and leads to series of dotted curves in Fig. \ref{scaling}. As expected
these curves are close to the solid ones for $P^2\ll M^2$, but
substantially different for $P^2$ approaching $M^2$ from below.
As emphasized at the beginning, PDF of virtual photons make good sense
for $P^2\ll M^2$ only and Fig. \ref{scaling} suggests that this means,
roughly, $P^2\lesssim 0.2 M^2$.
On the other hand, from the point of view of mathematics the evolution
equation (\ref{NSevolution}) as well as its solution (\ref{result})
can be considered for arbitrary $P^2$ and $M^2$.

In analyzing higher order corrections to (\ref{result})
I will make use of the following approximation
\begin{equation}
V(x_0)\equiv\int^{\infty}_{x_0}\frac{{\mathrm{d}}x}{x^2}
\frac{1}{\ln(x)}\doteq\frac{1}{x_0}\frac{1}{\ln(2x_0)},
\label{approximation}
\end{equation}
which for $2\le x_0\ll x_M$ is very accurate and allows us to
write
\begin{equation}
\frac{2P^2}{\beta_0}\int_{P^2}^{\infty}\frac{\mathrm{d}\tau}{\tau^2}
\frac{1}{\ln(\tau/\Lambda^2)}=\frac{2x_0}{\beta_0}\int_{x_0}^{\infty}
\frac{{\mathrm{d}}x}{x^2}\frac{1}{\ln(x)}
\doteq\frac{2}{\beta_0\ln(2x_0)}=
\frac{\alpha_s(2P^2)}{2\pi}=\frac{\alpha_s(P^2)}{2\pi}+
{\cal O}(\alpha_s^2(P^2))
\label{W}
\end{equation}
where $x\equiv \tau/\Lambda^2,\;x_0\equiv P^2/\Lambda^2$,
$x_M\equiv M^2/\Lambda^2$.
This integral first appears in evaluating the contribution of the
second part of the ${\cal O}(\alpha_s)$ term in (\ref{integrals}),
which has been left out in deriving (\ref{result})
\begin{equation}
-\int^1_x\frac{{\mathrm d}y}{y}P_{qq}^{(0)}\left(\frac{x}{y}\right)
k_L(y)P^2\int^{M^2}_{P^2}\frac{{\mathrm d}\tau_2}{\tau^2_2}
\frac{\alpha_s(\tau_2)}{2\pi}\doteq -\frac{\alpha_s(P^2)}{2\pi}
k_L\otimes P_{qq}^{(0)} .
\label{secondterm}
\end{equation}
In momentum space ({\ref{secondterm}) turns into
\begin{equation}
-\frac{\alpha_s(P^2)}{2\pi}k_L(n)P^{(0)}_{qq}(n)
\Rightarrow -\frac{\alpha_s(P^2)}{2\pi}k_L(n)P^{(0)}_{qq}(n)
\left[\frac{\alpha_s(M^2)}{\alpha_s(P^2)}\right]
^{-2P^{(0)}_{qq}(n)/\beta_0},
\label{secondmom}
\end{equation}
where the expression after the arrow follows when in all terms
on the r.h.s. of (\ref{integrals}) the inner integral over
${\mathrm d}\tau$ is approximated using (\ref{approximation}) as
$(\alpha_s(P^2)/P^2-\alpha_s(\tau_1)/\tau_1)$
and only the first term retained.
Starting from the third term on the r.h.s. of (\ref{integrals})
the second term, i.e. $\alpha_s(\tau_1)/\tau_1$ is not negligible
as the upper limit $\tau_1$ does not coincide with $M^2$, but spans
the whole range $P^2\le \tau_1\le M^2$. Taking it also into account
yields (\ref{result}) multiplied by
a term proportional to $\alpha_s^2(P^2)$. Repeating this
procedure at higher orders leads to the general expression
\begin{equation}
q_L^{\mathrm{NS}}(n,P^2,M^2)=k_L(n)
\left(1+\sum_{k=1}^{\infty}c_k(n)\alpha^k_s(P^2)\right)
\left[\frac{\alpha_s(M^2)}{\alpha_s(P^2)}\right]
^{-2P^{(0)}_{qq}(n)/\beta_0},
\label{allorders}
\end{equation}
where $c_1=-P^{(0)}_{qq}(n)/2\pi$ etc. Although calculable
all the coefficients $c_k(n),k\ge 1$ were mentioned only to indicate
the origin of the general structure of the results after the
integrations in (\ref{integrals}), but should be discarded in the
leading--log approximation, used in deriving (\ref{result}).

To estimate the hadronic part of $q_L^{\mathrm{NS}}$,
I have assumed that its onset is governed by the same ratio $P^2/m_V^2$
of the photon virtuality $P^2$ and vector meson mass $m_V^2$ as the
decrease of $q_T^{\mathrm{NS}}$ with increasing $P^2$:
\begin{equation}
q_L^{\mathrm{NS,HAD}}(x,P^2,M^2)\equiv r(P^2)q_T^{\mathrm{NS,HAD}}
(x,P^2,M^2),~~~r(P^2)\equiv \left(\frac{P^2}{P^2+m_V^2}\right).
\label{rescaled}
\end{equation}
In Fig. \ref{scaling}b (\ref{result}) is compared to (\ref{rescaled})
assuming $q_T^{\mathrm{NS,HAD}}=xu(x)/3e_u^2=3xu(x)$ and using the
SAS1D parameterization of $u(x,P^2,M^2)$
\footnote{In this way the hadronic part of $q_T^{\mathrm{NS}}$ is
in fact overestimated. For $P^2\gtrsim 2$ GeV$^2$ the suppression factor
$P^2/(P^2+m_V^2)$ is close to unity and has therefore only
small effect.}.
Estimated in such a way, the hadronic part of $q_L^{\mathrm{NS}}$
seems to be safely negligible for $P^2\gtrsim 1-2$ GeV$^2$.

\section{Pointlike part of $\gamma_L^*$ in QCD -- singlet and gluon
distributions}
The fact that $q_L^{\mathrm{NS}}$ satisfies homogeneous evolution
equation suggests the same for
the quark singlet and gluon distribution functions. Indeed, following
the procedure of previous Section one can resum the series coming from
diagrams in Fig. \ref{figgl} as follows
\begin{equation}
G_L(n,P^2,M^2)=\frac{2n_f\langle e^2\rangle
k_L(n)P^{(0)}_{Gq}(n)}{P^{(0)}_{GG}(n)}
\left(\left[\frac{\alpha_s(M^2)}{\alpha_s(P^2)}\right]^
{-2P^{(0)}_{GG}(n)/\beta_0}-1\right)=2n_f\langle e^2\rangle
G_L^{\mathrm{S}}.
\label{gluons1}
\end{equation}
\begin{figure}[t]\centering
\epsfig{file=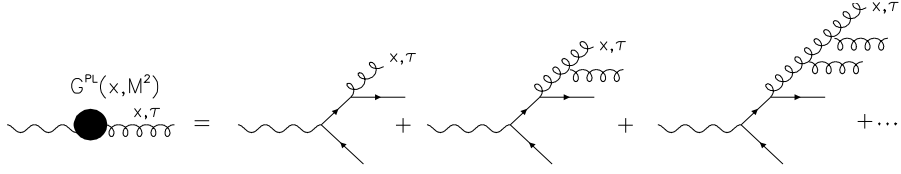,width=12cm}
\caption{Part of the set of diagrams defining the gluon
distribution function of the photon.}
\label{figgl}
\end{figure}
Adding multiple gluon emissions on the outgoing quark line
modifies (\ref{gluons1}) as follows
\begin{equation}
G_L^{\mathrm{S}}(n,P^2,M^2)=
\frac{k_L(n)P^{(0)}_{Gq}(n)}{P^{(0)}_{GG}(n)-
P^{(0)}_{qq}(n)}
\left(\left[\frac{\alpha_s(M^2)}{\alpha_s(P^2)}\right]^
{-2P^{(0)}_{GG}(n)/\beta_0}-
\left[\frac{\alpha_s(M^2)}{\alpha_s(P^2)}\right]^
{-2P^{(0)}_{qq}(n)/\beta_0}\right).
\label{gluons2}
\end{equation}
Combined with $q_L^{\mathrm{NS}}$ defined in (\ref{result})
$G_L^{\mathrm{S}}(n,P^,M^2)$ satisfies the evolution equation
\begin{equation}
\frac{{\mathrm{d}}G_L^{\mathrm{S}}(n,P^2,M^2)}{{\mathrm{d}}\ln M^2}=
\frac{\alpha_s(M}{2\pi}
\left[P^{(0)}_{GG}(n)G_L^{\mathrm{S}}(n,P^2,M^2)+
P^{(0)}_{Gq}(n)q_L^{\mathrm{NS}}(n,P^2,M^2)\right].
\label{gluonevolution}
\end{equation}
As for $q^{\mathrm{NS}}_L$ (\ref{gluons2}) was derived assuming
$P^2\ll M^2$, but from the point of view of mathematics can be
considered for arbitrary values of $P^2\le M^2$.
Note that whereas
$q_L^{\mathrm{NS}}(n,P^2,P^2)=k_L(n)$,
$G_L^{\mathrm{S}}(n,P^2,M^2)$ defined in
(\ref{gluons2}) vanishes for $P^2=M^2$.
\begin{figure}[h]\centering
\epsfig{file=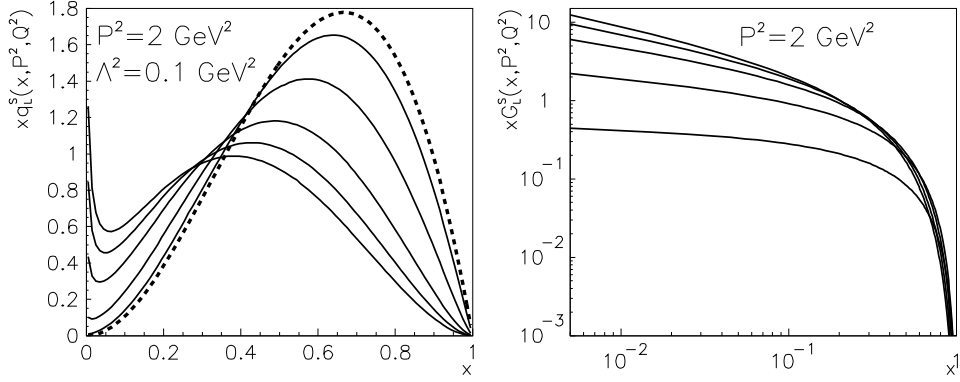,width=13cm}
\caption{Scale dependence of $q_L^{\mathrm{S}}$ and $G_L^{\mathrm{S}}$
for $n_f=4$ and
$\Lambda^2=0.1$ GeV$^2$. Ordered from below at small $x$ the curves
correspond to $M^2=3,10,10^2,10^3$ and $10^4$ GeV$^2$ respectively.}
\label{figsinglet}
\end{figure}
The equation (\ref{gluonevolution}) is not exactly the evolution
equation for gluons in hadrons as the latter contains quark singlet
instead, but the diagrams included in the definition (\ref{gluons2})
do not exhaust all possible ladder diagrams defining $G_L^{\mathrm{S}}$.
The above considerations are, however, sufficient to conclude that
``dressed'' PDF of $\gamma_L^*$ are constructed from the ``bare'' ones
in the same way as those of hadrons and therefore satisfy
the same evolution equations as the latter, with initial conditions at
$M^2=P^2$ having the form
\begin{equation}
q_L^{\mathrm{NS}}(x,P^2,P^2)=q_L^{\mathrm{S}}(x)=k_L(x)=12x(1-x),
~~~G_L^{\mathrm{S}}(x,P^2,P^2)=0.
\label{initial}
\end{equation}
Using the formulae of \cite{Buras} we can write explicitly
\begin{eqnarray}
q_L^{\mathrm{S}} &\!\!\! = &\!\!\!k_L(n)
\left[\left[\frac{\alpha_s(M^2)}{\alpha_s(P^2)}\right]^{d_+(n)}+
a(n)\left(\left[\frac{\alpha_s(M^2)}{\alpha_s(P^2)}\right]^{d_-(n)}-
\left[\frac{\alpha_s(M^2)}{\alpha_s(P^2)}\right]^{d_+(n)}\right)
\right],\label{qLL}\\
G_L^{\mathrm{S}} &\!\!\! = &\!\!\!k_L(n)\epsilon(n)
\left(\left[\frac{\alpha_s(M^2)}{\alpha_s(P^2)}\right]^{d_-(n)}-
\left[\frac{\alpha_s(M^2)}{\alpha_s(P^2)}\right]^{d_+(n)}\right),
\label{GLL}
\end{eqnarray}
\begin{eqnarray}
\kappa(n)& = & \sqrt{\left(P^{(0)}_{qq}(n)-P^{(0)}_{GG}(n)\right)^2+
8n_fP^{(0)}_{Gq}(n)P^{(0)}_{qG}(n)},\label{kappan}\\
a(n) & = & \frac{P^{(0)}_{qq}(n)-P^{(0)}_{GG}(n)+\kappa(n)}
{2\kappa(n)},~~~\epsilon(n)=\frac{P^{(0)}_{Gq}(n)}{\kappa(n)},
\label{epsilonn}\\
d_{\pm}(m) & = & \frac{-P^{(0)}_{qq}(n)-P^{(0)}_{GG}(n)\pm
\kappa(n)}{\beta_0}.\label{apm}
\end{eqnarray}
Together with $q_L^{\mathrm{NS}}$ as given in (\ref{result}) this
completes the evaluation of PDF of $\gamma_L^*$. The singlet charge
factor $2n_f\langle e^2\rangle$ factorizes  out of both $\Sigma_L$
and $G_L$  due to vanishing of $G_L$ at the initial scale $P^2$.
The nontrivial dependence on $n_f$ remains inside $\kappa(n)$.

Performing numerically the inverse Mellin transformation
of (\ref{qLL}-\ref{GLL}) leads the scale dependence of
$q_L^{\mathrm{S}}$ and $G_L^{\mathrm{S}}$ in $x$-space,
illustrated in Fig. \ref{figsinglet} for $P^2=2$ GeV$^2$,
$\Lambda^2=0.1$ GeV$^2$ and $n_f=4$.
The resulting parameterization
\footnote{Obtainable upon request from chyla@fzu.cz.}
of PDF of $\gamma_L^*$ based on eqs. (\ref{result}), (\ref{qLL}) and
(\ref{GLL}) contains $\Lambda$ as a free parameter. The fact that PDF
of $\gamma_L^*$ satisfy the homogeneous evolution equations implies
the validity of standard momentum sum rule. Recall that its violation
for $\gamma_T^*$ is due to the fact that the ``bare'' quark
distribution function of $\gamma_T^*$,
$q_T^{\mathrm{QED}}\propto\ln M^2$, depends on the factorization
scale $M$.

The QCD induced scale dependence of PDF of $\gamma_L^*$ is formally
reminiscent of the idea of ``dynamically generated'' PDF of hadrons
advocated by the GRV group (see \cite{GRV1} and references therein).
Note, however, that the initial conditions (\ref{initial}) themselves
have no direct physical interpretation as all our formulae were
derived under the assumption $P^2\ll M^2$.

\section{Phenomenological consequences}
In \cite{long} we discussed the numerical relevance of the contribution
of resolved $\gamma_L^*$ for two physical quantities:
$F_2^{\gamma}(x,P^2,M^2)$ and the {\em effective} PDF
\begin{equation}
D_{\mathrm{eff}}(x,P^2,M^2)\equiv \sum_{i=1}^{n_f}\left(q_(x,P^2,M^2)
+\overline{q}_i(x,P^2,M^2)\right)+\frac{9}{4}G(x,P^2,M^2)
\label{Deff}
\end{equation}
relevant for dijet production in ep and e$^+$e$^-$ collisions. The
contributions of $\gamma_L^*$ plotted in Figs. 2 and 3 of \cite{long} and
reproduced by dashed curves in Figs. \ref{eff2} below
were calculated using simple QED formula (\ref{QEDL}) for the quark
content of $\gamma_L^*$.
\begin{figure}\unitlength=1mm
\begin{picture}(160,95)
\put(20,47){\epsfig{file=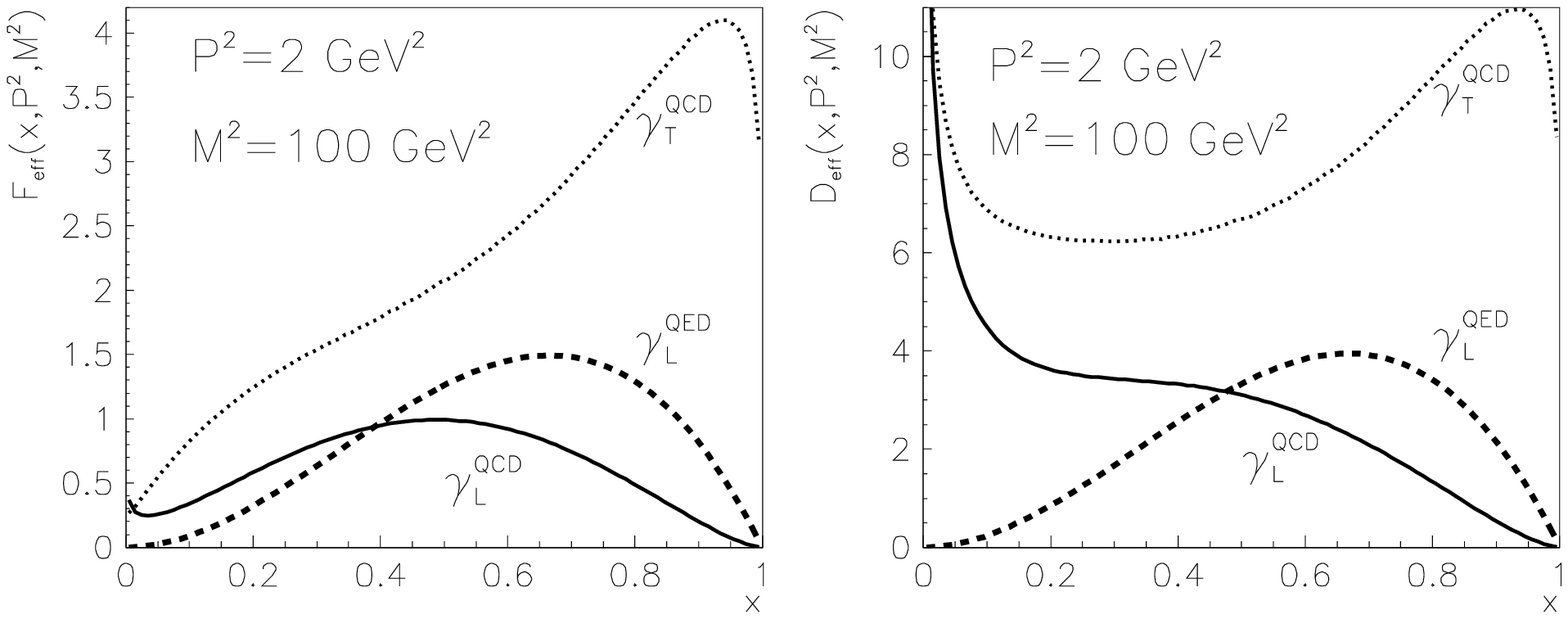,width=12cm}}
\put(20,-2){\epsfig{file=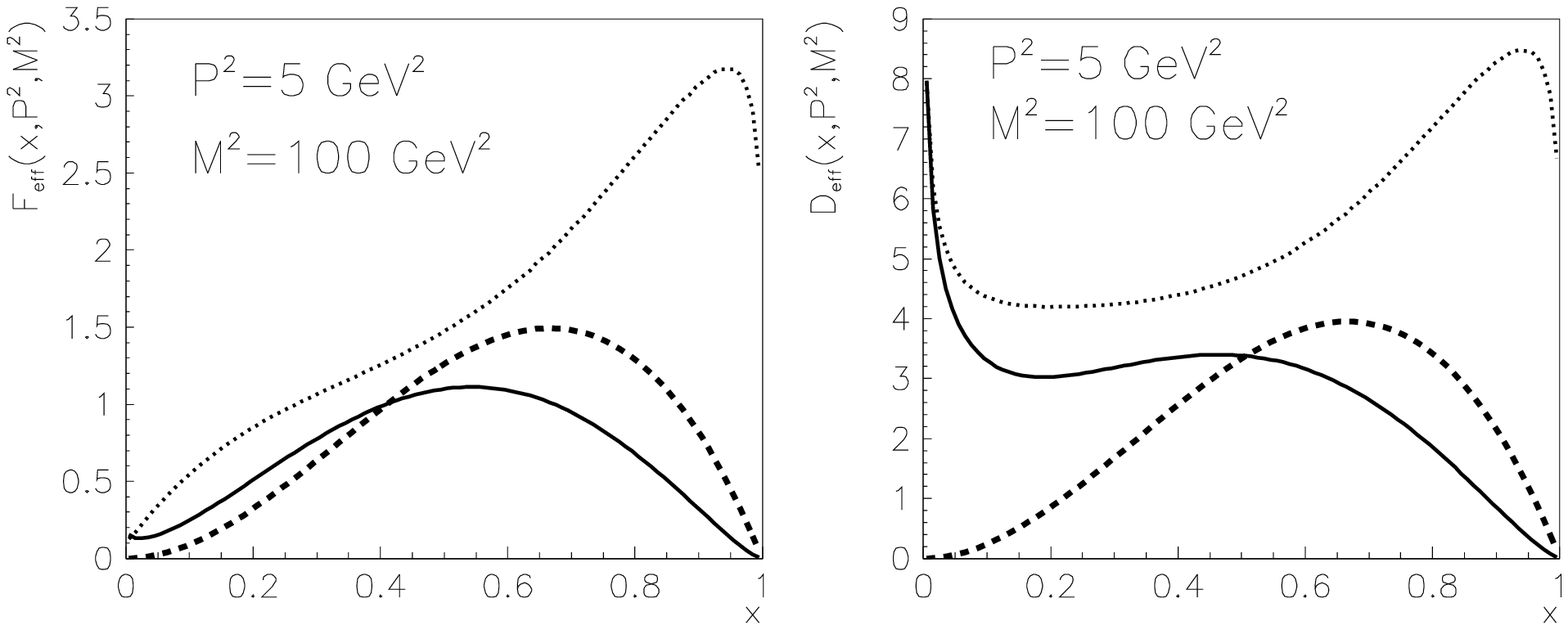,width=12cm}}
\end{picture}
\caption{Comparison of the contributions $F_{TT}$ and $F_{TL}$ of
target $\gamma_T^*$ and $\gamma_L^*$ to $F_{\mathrm{eff}}$
(left) and $D_{\mathrm{eff}}$ (right) for $P^2=2,5$ GeV$^2$
and $M^2=100$ GeV$^2$ as given by QED and QCD formulae (\ref{QEDL})
and (\ref{qLL}) for $\gamma_L^*$ and SAS1D parameterization for
$\gamma_T^*$.}
\label{eff2}
\end{figure}
They peaked around $x\simeq 0.7$ and were more pronounced for
$F_2^{\gamma}$ than for $D_{\mathrm{eff}}$. QCD corrections to PDF
of $\gamma_L^*$, given in eqs. (\ref{result}),(\ref{qLL}) and
(\ref{GLL}), modify this situation significantly. Compared
to purely QED predictions, the QCD effects suppress quark
distribution functions of $\gamma_L^*$ at large $x$ and enhance
them on the other hand for $x\lesssim 0.4$. This enhancement with
respect to the contribution of $\gamma_T^*$ is in fact so large that
for $x\lesssim 0.02$ the contribution of $\gamma_L^*$ to $F_2^{\gamma}$
exceeds that of $\gamma_T^*$. For $D_{\mathrm{eff}}$ the region
$x\lesssim 0.4$ is dominated by gluons from both $\gamma_T^*$ and
$\gamma_L^*$, and again the contributions of $\gamma_L^*$ is
comparable to those of $\gamma_T^*$. The relevance of $\gamma_L^*$
is further enhanced with increasing $P^2$ (but keeping $P^2\ll M^2$).
but decrease with increasing $M^2$. Taking into
account experimental conditions at
LEP and HERA, Figure \ref{eff2} suggests that dijet production at
HERA \cite{phd} offers particularly suitable place to look for the
effects of $\gamma_L^*$. The quantity $D_{\mathrm{eff}}$ allows us
to express the relevant cross sections in a simple but only
approximate way. Work on incorporating QCD improved PDF of
$\gamma_L^*$ within the LO Monte-Carlo event generators as well
as NLO parton level calculations is in progress.

\end{document}